
\input phyzzx
{}~\hfill\vbox{\hbox{TIFR/TH/91-47}\hbox{October, 1991}}\break

\title{GAUGE AND GENERAL COORDINATE INVARIANCE IN NON-POLYNOMIAL CLOSED
STRING FIELD THEORY}

\author{Debashis Ghoshal and Ashoke Sen\foot{e-mail addresses:
\tt{GHOSHAL@TIFRVAX.BITNET, SEN@TIFRVAX.BITNET}}}

\address{Tata Institute of Fundamental Research, Homi Bhabha Road, Bombay
400005, India}

\abstract

An appropriate field configuration in non-polynomial closed string field
theory is shown to correspond to a general off-shell field configuration
in low energy effective field theory. A set of string field theoretic
symmetries that act on the fields in low energy effective field theory as
general coordinate transformation and antisymmetric tensor gauge
transformation is identified.
The analysis is carried out to first order in the fields; thus the
symmetry transformations in string field theory reproduce the linear and
the first non-linear terms in the gauge transformations in the low energy
effective field theory.

\NPrefs
\def\define#1#2\par{\def#1{\Ref#1{#2}\edef#1{\noexpand\refmark{#1}}}}
\def\con#1#2\noc{\let\?=\Ref\let\<=\refmark\let\Ref=\REFS
         \let\refmark=\undefined#1\let\Ref=\REFSCON#2
         \let\Ref=\?\let\refmark=\<\refsend}

\define\SIEGELBOOK
W. Siegel, Introduction to String Field Theory, World Scientific, 1988.

\define\SCHUBERT
C. Schubert, MIT preprint CTP 1977.

\define\LPP
A. LeClair, M. Peskin and C. Preitschopf, Nucl. Phys. {\bf B317} (1989)
411,464.

\define\BPZ
A.~Belavin, A.~M.~Polyakov and A.~B.~Zamolodchikov, Nucl.~Phys. {\bf B241}
(1984) 333.

\define\SENPOLY
A.~Sen, Phys.~Lett. {\bf B241} (1990) 350,  Nucl.~Phys. {\bf B345} (1990)
551.

\define\KAKU
M. Kaku and J. Lykken, Phys. Rev. {\bf D38} (1988) 3067;
M. Kaku, preprints CCNY-HEP-89-6, Osaka-OU-HET 121.

\define\NONPOL
M.~Saadi and B.~Zwiebach, Ann.~Phys. {\bf 192} (1989) 213;
T.~Kugo, H.~Kunitomo, and K.~Suehiro, Phys.~Lett. {\bf 226B} (1989) 48;

\define\GAUGEINV
T.~Kugo and K.~Suehiro, Nucl.~Phys. {\bf B337} (1990) 434.

\define\SYCL
S.~Mukherji and A.~Sen,  Nucl. Phys. {\bf B363} (1991) 639.

\define\SOZW
H. Sonoda and B. Zwiebach, Nucl. Phys. {\bf B331} (1990) 592.

\define\HLOOP
H.~Hata, Phys.~Lett. {\bf 217B} (1989) 438, 445; Nucl.~Phys. {\bf B329}
(1990) 698; {\bf B339} (1990) 663.

\define\SAADI
M. Saadi, Mod. Phys. Lett. {\bf A5} (551) 1990; Int. J. Mod. Phys. {\bf
A6} (1991) 1003.

\define\ZWIEBACH
B.~Zwiebach, Mod.~Phys.~Lett. {\bf A5} (1990) 2753;
Phys. Lett. {\bf B241} (1990) 343; Comm. Math. Phys. {\bf 136} (1991) 83.

\define\MKAKU
L. Hua and M. Kaku, Phys. Lett. {\bf B250} (1990) 56;
M. Kaku, Phys. Lett. {\bf B250} (1990) 64.

\define\PREGEO
H. Hata and M. Maeno, preprint KUNS-1037 (1990).

\define\ZWIE
B. Zwiebach, Phys. Lett. {\bf B156} (1985) 315

\define\ZWIEPRI
B. Zwiebach, as quoted in ref.\SCHUBERT.

\define\SCHERK
J. Scherk and J. Schwarz, Nucl. Phys. {\bf B81} (1974) 118, Phys. Lett.
{\bf B52} (1974) 347.

\define\YONEYA
T. Yoneya, Nuovo Cimento Lett. {\bf 8} (1973) 951; Prog. Theor. Phys. {\bf
51} (1974) 1907; {\bf 56} (1976) 1310.

\def\cpa{\Phi_{1,\alpha}}
\def\psmj{\Psi^{(m)}}
\def\psmp{\Psi^{(m')}}
\def\lr{\Lambda^{(\rho)}}
\def\lrp{\Lambda^{(\rho')}}
\def\c{\circ}
\def\hpsm{\hat\Psi^{(\hat m)}}
\def\hlr{\hat\Lambda^{(\hat\rho)}}
\def\ck{{\cal K}}
\def\prc{\Phi^c_{3,r}}
\def\hprc{\hat\Phi^c_{3,{\hat r}}}
\def\tprc{\tilde\Phi^c_{3,{\tilde r}}}
\def\erk{\eta^{(\rho)}_\kappa}
\def\erpkp{\eta^{(\rho')}_{\kappa'}}
\def\herk{\hat\eta^{(\hat\rho)}_\kappa}
\def\terk{\tilde\eta^{(\tilde\rho)}_\kappa}
\def\pmi{\phi^{(m)}_i}
\def\pmj{\phi^{(m)}_j}
\def\pmpjp{\phi^{(m')}_{j'}}
\def\hpmj{\hat\phi^{(\hat m)}_j}
\def\tpmj{\tilde\phi^{(\tilde m)}_j}
\def\half{{1\over 2}}
\def\th{\tilde h}
\def\tb{\tilde b}
\def\tF{\tilde F}
\def\tP{\tilde P}
\def\tS{\tilde S}
\def\tE{\tilde E}
\def\co{{\cal O}}
\def\eps{\epsilon}
\def\teps{\tilde\eps}
\def\txi{\tilde\xi}
\def\heps{\hat\eps}
\def\hxi{\hat\xi}
\def\ka{\kappa}

\def\tA{\tilde A}
\def\ch{{\cal H}}
\def\p{\partial}
\def\to{\rightarrow}
\def\l{\langle}
\def\r{\rangle}
\def\tatw{\tA^{(2)}}

\def\atw{A^{(2)}}
\def\ath{A^{(3)}}
\def\btw{B^{(2)}}
\def\bth{B^{(3)}}
\def\ctw{C^{(0)}}
\def\cth{C^{(1)}}
\def\dtw{D^{(0)}}
\def\dth{D^{(1)}}
\def\kth{K^{(1)}}

\chapter{INTRODUCTION}

It is by now well established that the low energy effective field theory
describing any critical string theory contains gravity\SCHERK\YONEYA.
The chain of results which lead to this conclusion is the following.
First of all, one finds that the spectrum of free string theory contains a
massless rank 2 symmetric tensor state; the physical states being in one
to one correspondence with the physical states associated with the
graviton field.
Secondly, one finds that amplitudes involving this massless tensor state
satisfy the same set of Ward identities as the ones satisfied by
scattering amplitudes involving external graviton states in a generally
covariant theory.
Finally, the tree level amplitudes involving the massless tensor states
agree, at low energies, to the tree level amplitudes calculated from
Einstein's action.

More recently, a complete covariant field theory for closed bosonic
strings has been constructed\NONPOL\GAUGEINV.
(See also ref.\KAKU.)
This field theory is characterized by an infinite parameter gauge
invariance\GAUGEINV\SCHUBERT, and an infinite number of fields.
Quantization of this theory has been carried out in
refs.\con\HLOOP\SOZW\SAADI\ZWIEBACH\MKAKU\noc.
One expects that general coordinate transformation will emerge as a
particular combination of the gauge transformations in string theory, and
the metric will be related to the infinite component string field through a
suitable functional relation.
For previous work on this subject see ref.\PREGEO (also see
ref.\SIEGELBOOK\ for a review of linearized gauge invariance in string
field theory).

In this paper we investigate this connection in detail.
We show that at least to first non-linear order, it is possible to
identify string field configurations which correspond to specific
(off-shell) configuration of the fields that appear in the low energy
effective field theory, namely the metric, the dilaton and the
antisymmetric tensor gauge field.
To this order we also identify specific symmetries of the string field
theory which can be identified as gauge transformations of the low energy
effective field theory. This contains general coordinate
transformation and antisymmetric tensor gauge transformation.
It turns out that the gauge symmetries of the low energy effective field
theory are obtained by a combination of gauge transformation in string
field theory and a `trivial' symmetry of the form:
$$
\delta\psi_r = K_{rs}(\{\psi\}) {\delta S\over\delta\psi_s}
\eqn\eoneone
$$
where $\psi_r$ are the string field components, $K_{rs}$ is an
antisymmetric matrix valued function of $\psi_r$, and $S$ is the string
field theory action.\foot{This symmetry is called `trivial' as it exists
for all theories. However it is perhaps worth
emphasizing that this is a genuine symmetry of the action.
}

The plan of the paper is as follows.
In Sect.~2 we explain our general approach to solve the problem of
identifying the field configurations and gauge symmetries in low energy
effective field theory to those in string field theory.
We allow for the most general functional relation between string
fields and the low energy fields, and also between the gauge
transformations in string field theory and those in low energy effective
field theory. We then derive constraints that must be satisfied in order
that the gauge transformations in the low energy effective field theory
are compatible with those in string field theory.
In sect.~3 we show how a solution to these constraint equations may be
obtained at the lowest level, so that the linearized gauge transformations
involving the massless fields in the low energy effective field theory
agree with those in string field theory.
In sect.~4 we analyze the set of constraint equations derived in sect.~2
to the next order.
We show that a solution to these set of equations can be obtained
provided a certain set of consistency conditions are satisfied by the
interaction
vertices in string field theory.
Appendix A contains a verification of the fact that these consistency
conditions are
indeed satisfied by the vertices of string field theory. This completes
the proof that to first non-linear
order, suitable functional relations between the fields and
gauge transformation parameters appearing in string field theory and those
appearing in low energy effective field theory may be found so that gauge
invariance in low energy effective field theory can be derived as a
consequence of the symmetries of string field theory.
We conclude in sect.~5 with some speculations and implications of our
result for the closure of the gauge algebra in string field theory.

\chapter{GENERAL FORMALISM}

Let $\ch$ be the Hilbert space of the combined matter ghost conformal
field theory describing first quantized string theory, and $\{|\Phi_{2,r}\r\}$
denote a basis of states in $\ch$ with ghost number 2 and annihilated by
$b_0^-$ and $L_0^-$.
(We shall work with the convention $c_0^\pm=(c_0\pm\bar c_0)/\sqrt 2$,
$b_0^\pm = (b_0\pm\bar b_0)/\sqrt 2$, and $L_0^\pm=(L_0\pm\bar L_0)/\sqrt
2$.)
Then the string field $|\Psi\r$ may be expanded as
$b_0^-|\Psi\r=\sum_r\psi_r|\Phi_{2,r}\r$ and the string field theory action is
given by\NONPOL\GAUGEINV\SENPOLY\
$$
S(\Psi) ={1\over 2}\l\Psi|Q_B b_0^-|\Psi\r +\sum_{N=3}^\infty
{g^{N-2}\over N!}\{\Psi^N\}
\equiv\sum_{N=2}^\infty {1\over N} \tA^{(N)}_{r_1\ldots r_N}\psi_{r_1}
\ldots \psi_{r_N},
\eqn\etwoone
$$
where $Q_B$ is the BRST charge of the first quantized string theory, and
$\{~~\}$ has been defined in refs.\GAUGEINV\SENPOLY.
Throughout this paper we shall use the convention of ref.\SENPOLY.
The coefficients $\tA^{(N)}$ are symmetric in all the indices
and are given by,
$$
\eqalign{
\tA^{(2)}_{r_1r_2} &= -\l\Phi_{2,r_1}|c_0^- Q_B|\Phi_{2,r_2}\r,\cr
\tA^{(N)}_{r_1\cdots r_N} &= {g^{N-2}\over (N-1)!} \{(c_0^-\Phi_{2,r_1})\cdots
     (c_0^-\Phi_{2,r_N})\},\qquad N\ge 3.\cr}
\eqn\etwonea
$$
Let $|\cpa\r$ denote a basis of states in $\ch$ of ghost number 1
and annihilated by $b_0^-$ and $L_0^-$.
Then the gauge transformation parameter $|\Lambda\r$ in string field
theory may be expanded as
$b_0^-|\Lambda\r=\sum_\alpha\lambda_\alpha|\cpa\r$, and the gauge
transformation in string field theory takes the form:
$$
\delta (b_0^-|\Psi\r) = Q_B b_0^-|\Lambda\r+\sum_{N=3}^\infty
{g^{N-2}\over (N-2)!}[\Psi^{N-2}\Lambda]
\eqn\etwotwo
$$
with $[~~]$ as defined in refs.\GAUGEINV\SENPOLY.
Let us introduce a basis of bra states $\l\Phi_{3,r}^c|$ of ghost number 3,
annihilated by $b_0^-$ and $L_0^-$, and satisfying,
$$
\l\Phi^c_{3,r}|c_0^-|\Phi_{2,s}\r =\delta_{rs}
\eqn\efourfive
$$
so that $\{\l\Phi^c_{3, r}|\}$ form a basis conjugate to
$\{|\Phi_{2,s}\r\}$. We can now write eq.~\etwotwo\ in
terms of the component fields as follows:
$$
\delta\psi_s =\sum_{N=2}^\infty A^{(N)}_{s\alpha r_1\ldots
r_{N-2}}\lambda_\alpha\psi_{r_1}\ldots\psi_{r_{N-2}},
\eqn\etwothree
$$
where,
$$
\eqalign{
A^{(2)}_{s\alpha} &= \l\Phi^c_{3,s}|c_0^- Q_B|\Phi_{1,\alpha}\r,\cr
A^{(N)}_{s\alpha r_1\cdots r_{N-2}} &={g^{N-2}\over (N-2)!}
\l\Phi^c_{3,s}| c_0^- [(c_0^-\Phi_{1,\alpha})
     (c_0^-\Phi_{2,r_1})\cdots (c_0^-\Phi_{2,r_{N-2}})]\r\cr
&= -{g^{N-2}\over (N-2)!}
     \{(c_0^-\Phi^c_{3,s}) (c_0^-\Phi_{1,\alpha})
     (c_0^-\Phi_{2,r_1})\cdots (c_0^-\Phi_{2,r_{N-2}})
     \},\qquad N\ge 3.\cr}
\eqn\etwothreea
$$
Note that $A^{(N)}_{s\alpha r_1\cdots r_{N-2}}$ is symmetric in the
indices $r_1, r_2,\cdots,r_{N-2}$.

Closed bosonic string theory at low energies is described by the
effective field theory involving the
graviton, the dilaton, and the antisymmetric tensor gauge fields.
Let $\{\phi_i\}$ denote the set of all the dynamical degrees of freedom of
the low energy effective field theory, where the index $i$ stands for the
field index, as well as space-time coordinates (or, equivalently, momenta
if we are working with Fourier transforms of the fields).
Also, let $\{\eta_\ka\}$ denote the set of gauge transformation parameters
of the low energy effective field theory, where the index $\ka$ again
includes coordinate (momentum) index. The set $\{\eta_\ka\}$
contains general coordinate transformation and antisymmetric tensor
gauge transformation.
Let the general form of the gauge transformation in the low energy
effective field theory be:
$$
\delta\phi_i=\sum_{N=2}^\infty B^{(N)}_{i j_1\ldots j_{N-2}\ka}
\eta_\ka\phi_{j_1} \ldots \phi_{j_{N-2}}
\eqn\etwofour
$$
The coefficients $B^{(N)}$ are all known from the low energy
effective field theory.
We now ask the following question: Can this low energy effective field
theory
be obtained (after some possible field redefinition) by integrating
out the massive modes of the string field theory?
If so, then any off shell configuration described by some arbitrary
choice of the variables $\phi_i$ should correspond to some configuration
of the string fields $\psi_r$.
(This configuration certainly involves massive modes of the string.
However the values they take are specified by their equations of motion.)
We should also note that the configuration may not be unique --- a given
off shell configuration
of the low energy effective field theory may correspond to more than one
string field configurations which are related to each other by some gauge
symmetry associated with higher level states (or by gauge symmetries at the
massless level that have no counterpart in the effective field theory).
Let us call $\psi_r(\phi_i)$ to be one of the string field configurations
that correspond to a given configuration of the low energy fields $\phi_i$,
and allow for the most general form of this function:
$$
\psi_r(\phi_i)=\sum_{N=0}^\infty {1\over (N+1)} C^{(N)}_{r i_1\ldots
i_{N+1}} \phi_{i_1}\ldots\phi_{i_{N+1}}
\eqn\etwofive
$$
The gauge transformation \etwofour\ of $\phi_i$ induces a
transformation on $\psi_r$.
The first few terms take the form:
$$\eqalign{
\delta\psi_r(\phi_i)=&\ctw_{ri}\btw_{i\ka}\eta_\ka +\ctw_{ri}\bth_{ij\ka}
\phi_j\eta_\ka
+\cth_{rij}\phi_i\btw_{j\ka}\eta_\ka +\co (\phi^2)\cr
}
\eqn\etwosix
$$
The question we are interested in can be formulated as follows.
Can we identify a gauge transformation parameter
$b_0^-|\Lambda(\eta,\phi)\r
=\sum_\alpha \lambda_\alpha(\eta,\phi)|\cpa\r$ in string field
theory such that the transformation \etwosix\ may be regarded as a gauge
transformation of the string field with this parameter?
If the answer is in the affirmative, then we could say that we have been
able to identify the off-shell gauge transformations of low energy
effective field theory to specific gauge transformations in string field
theory.
Again, we shall allow for the most general dependence of $\lambda_\alpha$
on $\phi_i$ and $\eta_\ka$ (keeping terms linear in $\eta_\ka$ only,
since we are considering infinitesimal gauge transformation):
$$
\lambda_\alpha(\eta,\phi) = \sum_{N=0}^\infty D^{(N)}_{\alpha\ka i_1\ldots
i_N}
\eta_\ka \phi_{i_1}\ldots \phi_{i_N}
\eqn\etwoseven
$$
Using eqs.\etwothree, \etwoseven\ and \etwofive\ we get,
$$\eqalign{
\delta\psi_r =&\atw_{r\alpha}\dtw_{\alpha\ka}\eta_\ka + \atw_{r\alpha}
\dth_{\alpha\ka i}\phi_i\eta_\ka
+\ath_{r\alpha s}\dtw_{\alpha\ka}\ctw_{s i}\phi_i\eta_\ka +
\co(\phi^2)\cr
}
\eqn\etwoeight
$$
Comparing eqs.\etwosix\ and \etwoeight\ we get,
$$\eqalign{
\ctw_{ri}\btw_{i\ka} = &\atw_{r\alpha}\dtw_{\alpha\ka}\cr
\ctw_{ri}\bth_{ij\ka} + \cth_{rji}\btw_{i\ka} = &
\atw_{r\alpha}\dth_{\alpha\ka j} + \ath_{r\alpha
s}\dtw_{\alpha\ka}\ctw_{sj} \cr
\cdots & \cr
\cdots & \cr
}
\eqn\etwonine
$$
Thus the question is, can we find appropriate $C^{(N)}$'s and $D^{(N)}$'s
such that the set of equations \etwonine\ are satisfied?
As we shall see, these equations cannot be satisfied beyond
lowest order, implying that off-shell general coordinate and
antisymmetric tensor
gauge transformations cannot be identified to pure gauge transformations
in string field theory.
Note, however, that besides gauge invariance, the string field theory
action is also trivially invariant under the transformation:
$$
\delta_{extra}\psi_r = K_{rs}(\{\psi_t\}) {\delta S\over\delta\psi_s}
\eqn\etwoten
$$
for any $K_{rs}$ which is antisymmetric in $r$ and $s$.
We shall show that a combination of gauge transformation in string field
theory given in eq.\etwothree\ and the transformation given in
eq.\etwoten\ can indeed be used to
generate general coordinate and antisymmetric tensor gauge transformations
in the low energy effective field theory, at least to first order in
$\psi_r$.
For this we allow for the most general choice of $K_{rs}$ as a function of
$\phi_i$ and $\eta_\ka$ (keeping terms linear in $\eta_\kappa$ only):
$$
K_{rs}=\sum_{N=1}^\infty K^{(N)}_{rs\ka i_1\ldots i_{N-1}} \eta_\ka
\phi_{i_1} \ldots \phi_{i_{N-1}}
\eqn\etwoeleven
$$
Using eqs.\etwoten, \etwoeleven, \etwoone\ and \etwofive\ we get:
$$
\delta_{extra}\psi_r =\kth_{rs\ka}\tatw_{st}\ctw_{ti}\phi_i\eta_\ka
+\co(\phi^2)
\eqn\etwotwelve
$$
Using eqs.\etwoeight\ and \etwotwelve\ we get the net transformation of
$\psi_r$ as,
$$\eqalign{
\delta_{tot}\psi_r = &\atw_{r\alpha}\dtw_{\alpha\ka}\eta_\ka
+ (\atw_{r\alpha}\dth_{\alpha\ka i} + \ath_{r\alpha s} \dtw_{\alpha\ka}
\ctw_{si} + \kth_{rs\ka} \tatw_{st}\ctw_{ti})\phi_i\eta_\ka + \co(\phi^2)
\cr
}
\eqn\etwothirteen
$$
Comparing eqs.\etwosix\ and \etwothirteen\ we get,
$$\eqalign{
\ctw_{ri}\btw_{i\ka} = & \atw_{r\alpha}\dtw_{\alpha\ka}\cr
\ctw_{ri}\bth_{ij\ka} + \cth_{rji} \btw_{i\ka} = & \atw_{r\alpha}
\dth_{\alpha\ka j} + \ath_{r\alpha s}\dtw_{\alpha\ka} \ctw_{sj} +
\kth_{rs\ka} \tatw_{st}\ctw_{tj}\cr
\cdots &\cr
\cdots &\cr
}
\eqn\etwofourteen
$$
Thus we now need to show that one can find appropriate $C^{(N)}$'s,
$D^{(N)}$'s and $K^{(N)}$'s so as to satisfy eq.\etwofourteen.
We shall show in the next two sections that this can be done at least for
the first and second equations in eqs.\etwofourteen.

\chapter{LINEARIZED OFF-SHELL GAUGE TRANSFORMATIONS}

In this section we shall obtain solution to the first of
eqs.\etwofourteen, thereby reproducing linearized general coordinate
invariance and antisymmetric tensor gauge invariance from string field
theory.
Although such analysis has been carried out in the past (see, for example
ref.\SIEGELBOOK), we shall repeat the analysis here for the sake of
completeness, and also to set up the notations that we shall be using in
the next section.

At the linearized level the metric $G_{\mu\nu}$, the antisymmetric tensor
field $B_{\mu\nu}$ and the dilaton $D$ appearing in the low energy
effective field theory transform as,
$$
\eqalign{\delta G_{\mu\nu} = & \p_\mu\eps_\nu + \p_\nu\eps_\mu\cr
\delta B_{\mu\nu} =& \p_\mu\xi_\nu -\p_\nu\xi_\mu\cr
\delta D =& 0\cr
}
\eqn\ethreeone
$$
where $\eps_\mu$ is the parameter labelling infinitesimal general
coordinate transformation and $\xi_\mu$ is the parameter labelling
infinitesimal antisymmetric tensor gauge transformation.
On the other hand, if we take a gauge transformation parameter in string
field theory
$$
b_0^-|\Lambda\r = \int d^D k[i\teps_\mu(k)(c_1\alpha^\mu_{-1} -\bar
c_1\bar\alpha^\mu_{-1}) + i\txi_\mu(k)(c_1\alpha^\mu_{-1}+\bar
c_1\bar\alpha^\mu_{-1}) +\sqrt 2\txi c_0^+]|k\r
\eqn\ethreetwo
$$
and define the string field components at the massless level as,\foot{At
the linearized level we would expect an off-shell configuration in low
energy effective field theory to correspond to a string field
configuration in which only the massless modes are excited, but beyond the
linearized order all the modes will be present.}
$$\eqalign{
b_0^-|\Psi\r =& \int d^D k[\half\th_{\mu\nu}(k) c_1\bar
c_1(\alpha^\mu_{-1}
\bar\alpha^\nu_{-1} +\bar\alpha^{\mu}_{-1}\alpha^\nu_{-1})
+i\sqrt 2 \tP_\mu(k) c_0^+(c_1\alpha^\mu_{-1} -\bar
c_1\bar\alpha^\mu_{-1}) \cr
&\qquad\qquad - \tF(k)(c_1 c_{-1} -\bar c_1\bar c_{-1}) +\half
\tb_{\mu\nu}(k) c_1\bar
c_1 (\alpha^\mu_{-1}\bar\alpha^\nu_{-1}
-\bar\alpha^\mu_{-1}\alpha^\nu_{-1}) \cr
&\qquad\qquad +i\sqrt 2\tS_\mu(k) c_0^+(c_1\alpha^\mu_{-1}+\bar
c_1\bar\alpha^\mu_{-1}) - \tilde E(k) (c_1 c_{-1} +\bar c_1\bar
c_{-1})]\>|k\r \cr
}
\eqn\ethreethree
$$
then the linearized gauge transformation $\delta (b_0^-|\Psi\r)= Q_B
b_0^-|\Lambda\r$ takes the form:
$$\eqalign{
\delta\th_{\mu\nu} =& -i(\teps_\mu k_\nu + k_\mu\teps_\nu), ~~~~~~
\delta\tP_\mu ={k^2\over 2}\teps_\mu,~~~~~~ \delta \tF =ik^\mu\teps_\mu\cr
\delta\tb_{\mu\nu} =& i(k_\mu\txi_\nu - k_\nu\txi_\mu), ~~~~~~
\delta\tS_\mu
= {k^2\over 2} \txi_\mu + ik_\mu\txi, ~~~~~~ \delta\tE(k) = i\txi_\mu
k^\mu - 2\txi\cr
}
\eqn\ethreefour
$$
Here $\alpha^\mu_{n}$ denote the modes of the world-sheet scalar fields
$X^\mu$.
Let us define,
$$\eqalign{
h_{\mu\nu} =&\th_{\mu\nu}\cr
P_\mu =& \tP_\mu -{i\over 2} k^\nu\th_{\mu\nu} -{i\over 2} k_\mu
\tF\cr
F =& \tF +\half h_\mu^\mu\cr
b_{\mu\nu} = &\tb_{\mu\nu}\cr
S_\mu =& \tS_{\mu} -{i\over 2} k^\nu\tb_{\mu\nu} +{i\over 2} k_\mu\tE\cr
E =&\tE\cr
}
\eqn\ethreefive
$$
for the fields; and similarly for the gauge parameters
$$\eqalign{
\heps_\mu =& -\teps_\mu\cr
\hxi_\mu =&\txi_\mu\cr
\hxi =&-{i\over 2}\txi_\mu k^\mu +\txi\cr
}
\eqn\ethreesix
$$
Transformation laws of various fields now take the form:
$$\eqalign{
\delta h_{\mu\nu} =&i(\heps_\mu k_\nu + \heps_\nu k_\mu)\cr
\delta P_\mu =& 0,\qquad\qquad \delta F = 0\cr
\delta b_{\mu\nu} =& i(k_\mu\hxi_\nu - k_\nu\hxi_\mu)\cr
\delta S_\mu =& 0,\qquad\qquad \delta E = -2\hxi\cr
}
\eqn\ethreeseven
$$
Comparing with eq.\ethreeone\ we see that at this level we can make the
identification:
$$\eqalign{
G_{\mu\nu}(x) = & \eta_{\mu\nu} - \sqrt 2 g \int d^D k h_{\mu\nu}(k)
e^{ik.x} \cr
B_{\mu\nu} (x) =& -\sqrt 2 g\int d^D k b_{\mu\nu}(k) e^{ik.x}\cr
D(x) =& -\sqrt 2 g\int d^D k F(k) e^{ik.x}\cr
}
\eqn\ethreeeight
$$
and,
$$\eqalign{
\eps_\mu(x) =& -\sqrt 2 g\int d^D k\heps_\mu(k) e^{ik.x}\cr
\xi_\mu(x) =& -\sqrt 2 g\int d^D k\hxi_\mu(k) e^{ik.x}\cr
}
\eqn\ethreenine
$$
The proportionality factor of $-\sqrt 2 g$ was worked out in ref.\SYCL.
(See also ref.\GAUGEINV.)

Note that there are three extra set of fields: $P_\mu$, $S_\mu$ and $E$,
and one extra gauge transformation parameter $\hxi$.
Eq.\ethreeseven\ shows that $E$ corresponds to a pure gauge deformation
generated by the parameter $\hxi$, hence we can set $E$ to be 0 by
adjusting $\hxi$.
This also removes the spurious gauge degrees of freedom associated with
the parameter $\hxi$.
The fields $P_\mu$ and $S_\mu$, on the other hand, can be identified as
auxiliary fields, as can be easily seen from the linearized equations of
motion $Q_B b_0^-|\Psi\r =0$.
The equations involving the fields $P_\mu$ and $S_\mu$ take the form:
$$
P_\mu =0,\qquad\qquad S_\mu =0
\eqn\ethreeten
$$
Since these equations are purely algebraic, we can set these fields to 0
to this order by using their equations of motion.
The remaining degrees of freedom are then in one to one correspondence
with the physical degrees of freedom of the low energy effective field
theory, and the remaining gauge transformation reproduces exactly the
linearized (but off-shell) gauge transformation of the low energy
effective field theory.

Using eqs.\ethreetwo, \ethreethree, \ethreefive\ and \ethreesix\ we get,
$$
\eqalignno{
b_0^-|\Lambda\r &= i\int d^D k\Big[-\heps_\mu(k)( c_1\alpha^\mu_{-1} -\bar
c_1\bar\alpha^\mu_{-1} )
+\hxi_\mu(k) (c_1\alpha^\mu_{-1} +\bar c_1
\bar\alpha^\mu_{-1} +{1\over\sqrt 2} k^\mu c_0^+)\Big]\>|k\r\cr
&&\eqname\ethreeeleven\cr
b_0^-|\Psi\r &= \int d^D k\Big[h_{\mu\nu}(k) c_1\bar c_1\alpha^\mu_{-1}
\bar\alpha^\nu_{-1}\cr
&\qquad\qquad +\sqrt 2 (-\half k^\nu h_{\nu\mu}(k) +{1\over 4}
k_\mu h_\nu^\nu (k) -\half k_\mu F(k)) c_0^+ (c_1\alpha^\mu_{-1} -\bar
c_1\bar\alpha^\mu_{-1}) \cr
&\qquad\qquad - (F(k) -\half h_\mu^\mu(k)) (c_1 c_{-1} -\bar c_1\bar
c_{-1}) +
b_{\mu\nu}(k) c_1\bar c_1 \alpha^\mu_{-1}\bar\alpha^\nu_{-1}\cr
&\qquad\qquad - {1\over\sqrt 2} k^\nu b_{\mu\nu}(k) c_0^+
(c_1\alpha^\mu_{-1} +\bar
c_1\bar\alpha^\mu_{-1})\Big] \>|k\r
&\eqname\ethreetwelve\cr
}
$$
where we have set $P_\mu$ and $S_\mu$ to 0 by their equations of motion,
and $E$ to zero by using the gauge invariance generated by $\hxi(k)$.
Using eqs.\ethreeeight\ and \ethreenine\ we see that $h_{\mu\nu}(k)$,
$b_{\mu\nu}(k)$ and $F(k)$ may be identified to the dynamical variables
$\phi_i$ that appear in the low energy effective action, whereas the
parameters $\heps_\mu(k)$ and $\hxi_\mu(k)$ may be identified to the gauge
transformation parameters $\eta_\ka$ that appear in the low energy
effective action.
Comparing with eqs.\etwofive\ and \etwoseven\ of sect.~2 we see that
eqs.\ethreeeleven\ and \ethreetwelve\ give the expressions for
$|\Lambda(\eta,\phi)\r$ and $|\Psi(\phi)\r$ to order $\phi^0$ and $\phi$
respectively such that the
gauge transformations in string field theory match those in the low energy
effective field theory.
In other words, it gives expressions for $\ctw_{ri}$ and
$\dtw_{\alpha\ka}$ satisfying the first of eqs.\etwofourteen:
$$
\ctw_{r i}\btw_{i\ka} = \atw_{r\alpha} \dtw_{\alpha\ka}
\eqn\ethreetwelvea
$$
Substituting eq.\ethreetwelve\ into the expression \etwoone\ for the
string field theory action we get, to quadratic order:
$$\eqalign{
S \propto & \int d^D k\bigl[h^{\mu\nu}(-k) ({k^2\over 2} h_{\mu\nu}(k) -
k_\mu
k^\sigma h_{\nu\sigma}(k) +\half k_\mu  k_\nu h_\sigma^\sigma (k) - k_\mu
k_\nu F(k)) \cr
&\qquad\qquad -2 \big(F(-k) -\half h_\mu^\mu(-k)\big) \big (k^2 F(k) -\half
(k^2\eta^{\rho\sigma} -
k^\rho k^\sigma) h_{\rho\sigma}(k)\big)\cr
&\qquad\qquad +b^{\mu\nu} (-k) (\half k^2 b_{\mu\nu}(k) - k_\nu k^\rho
b_{\mu\rho}(k))\bigr]\cr
}
\eqn\ethreethirteen
$$
which agrees with the low energy effective action.

Finally, note that the identification of the low energy fields from their
gauge transformation laws can be made only up to field redefinitions which
do not change the gauge transformation laws of various fields.
At the linearized level such field redefinitions will take the form
$h_{\mu\nu}\to h_{\mu\nu}+\Delta h_{\mu\nu}$, $F\to F+\Delta F$ and
$b_{\mu\nu}\to b_{\mu\nu}+\Delta b_{\mu\nu}$
where $\Delta
h_{\mu\nu}$, $\Delta F$ and $\Delta b_{\mu\nu}$ are gauge invariant linear
functions of $h_{\mu\nu}$, $b_{\mu\nu}$, $F$ and $k^\mu$.
In addition, if we want to express the action in a form such that the
quadratic terms in the action contain only two derivatives\ZWIE, then one
should be careful while adding momentum dependent terms in $\Delta
h_{\mu\nu}$, $\Delta b_{\mu\nu}$ and $\Delta F$.
In the present situation this leaves us with a field redefinition of the
form $h_{\mu\nu}\to h_{\mu\nu}+ a F\eta_{\mu\nu}$ where $a$ is an
arbitrary constant.
In the context of low energy effective field theory this corresponds to a
field redefinition of the form $G_{\mu\nu}\to f(D)G_{\mu\nu}$ where $f(D)$
is an arbitrary function of $D$.

\chapter{FIRST ORDER NON-LINEAR TERMS IN THE GAUGE TRANSFORMATION}

In this section we shall show that the second of eq.\etwofourteen,
$$
\ctw_{ri}\bth_{ij\ka} +\cth_{rji}\btw_{i\ka} =\atw_{r\alpha}
\dth_{\alpha\ka j} +\ath_{r\alpha s} \dtw_{\alpha\ka} \ctw_{sj}
+\kth_{rs\ka} \tatw_{st} \ctw_{tj}
\eqn\efourone
$$
can be satisfied by appropriately choosing $\cth_{rji}$, $\dth_{\alpha\ka
j}$ and $\kth_{rs\ka}$.

In analyzing these equations we shall first try to look for possible
obstructions to solving these equations.
Such obstructions are caused by equations which follow from eq.\efourone\
and are completely independent of $\cth_{rji}$, $\dth_{\alpha\ka j}$ and
$\kth_{rs\ka}$.
These would then correspond to equations involving known constants and
would have to be satisfied identically.
For this, let us choose a complete set of gauge transformations $\{\erk\}$
and a complete set of field configurations $\{\pmi\}$ in the low energy
effective field theory, and rewrite eq.\efourone\ as:
$$\eqalign{
\ctw_{ri}\bth_{ij\ka}\pmj\erk +\cth_{rji}\btw_{i\ka}\pmj\erk = &
\atw_{r\alpha} \dth_{\alpha\ka j}\pmj\erk + \ath_{r\alpha
s}\dtw_{\alpha\ka} \ctw_{sj}\pmj\erk\cr
& +\kth_{rs\ka}\tatw_{st}\ctw_{tj}\pmj\erk, \qquad {\rm for~all}~r, m,
\rho.\cr
}
\eqn\efourtwo
$$
We now divide the set $\{\pmj\}$ into two linearly independent sets,
$\{\hpmj\}$ and
$\{\tpmj\}$, satisfying,
$$
\eqalign{
\tatw_{st}\ctw_{tj}\hpmj &= 0, \qquad\hbox{for all }s;\cr
\tatw_{st}\ctw_{tj}\tpmj &\ne 0, \qquad\hbox{for some } s.\cr
}
\eqn\efourthree
$$
In other words $\hpmj$ denote field configurations which are solutions of the
linearized equations of motion, $\tpmj$ denote those which are not.
Similarly we divide the set $\{\erk\}$ into linearly independent sets
$\{\herk\}$ and $\{\terk\}$ satisfying:
$$
\eqalign{
\btw_{i\ka}\herk &= 0, \qquad\hbox{for all } i;\cr
\btw_{i\ka}\terk &\ne 0, \qquad\hbox{for some } i.\cr
}
\eqn\efourfour
$$
In other words, $\herk$ denote the set of gauge transformations for which
the field independent components in the expression for $\delta\phi_i$
vanishes.
$\{\herk\}$ thus includes rigid translation and rigid antisymmetric tensor
gauge transformations, as well as antisymmetric tensor gauge
transformations of the form $\xi_\mu=\p_\mu\chi$ for some
$\chi$.\foot{Note that global rotation is not included in the set
$\{\herk\}$, since the gauge transformation parameters for global rotation
blow up at infinity.
Another way of saying this is that the gauge transformation parameters are
linear in the space-time coordinates $X^\mu$, and hence the corresponding
states in $\ch$ do not correspond to well defined local fields in the
conformal field theory.}

Recall that $\{\l\Phi^c_{3, r}|\}$ form a basis conjugate to
$\{|\Phi_{2,s}\r\}$.
We now divide the set of states $\{\l\Phi^c_{3, r}|\}$ into two
sets $\{\l\hprc|\}$
and $\{\l\tprc|\}$ such that,
$$
\l\hprc| Q_B = 0,\qquad\quad\l\tprc| Q_B\ne 0.
\eqn\efoursix
$$
Let us now consider eq.\efourtwo\ with the free indices $m$, $\rho$ and
$r$ restricted to $\hat m$, $\hat \rho$ and $\hat r$ respectively.
Thus in this case the terms involving $\cth$ and $\kth$ vanish by
eq.\efourfour\ and \efourthree\ respectively.
Furthermore, from eqs.~\etwothreea\ and \efoursix\ it follows that,
$$
\atw_{\hat r\alpha} =0.
\eqn\efoureight
$$
In deriving the above equation we have used the fact that $\l\hprc|\co
|\Phi_{1,\alpha}\r =0$ if the operator $\co$ does not contain the mode
$c_0^-$.
Thus the term involving $\dth$ in eq.\efourtwo\ also vanishes when we
choose the index $r$ to be $\hat r$.
This equation may then be written as:
$$
\ctw_{\hat r i}\bth_{ij\ka}\hpmj\herk =\ath_{\hat r\alpha
s}\dtw_{\alpha\ka} \ctw_{sj}\hpmj\herk
\eqn\efournine
$$
Note that the undetermined coefficients $\cth$, $\dth$ and $\kth$
have disappeared from the above equation, whereas $\ctw$ and $\dtw$ have
already been determined in the previous section.
Hence unless the above
equation is satisfied identically, there is a genuine obstruction to
solving the set of equations \efourtwo.
We  show in appendix A that eq.\efournine\ is satisfied identically.
For the time being we assume this to be true and look for other possible
obstructions of this kind.

Next, let us restrict the index $r$ to be of type $\hat r$, and take
$\pmj$ to be of the form $\btw_{j\ka'}\erpkp$.
This causes the term involving $\dth$ in eq.\efourtwo\ to vanish, since
$\atw_{\hat r\alpha}=0$.
On the other hand, using eqs.\ethreetwelvea\ the term involving $\kth$ in
eq.\efourtwo\ may be brought to the form:
$$
\kth_{rs\ka}\tatw_{st}\ctw_{tj}\btw_{j\ka'}\erpkp\erk
=\kth_{rs\ka}\tatw_{st}\atw_{t\alpha}\dtw_{\alpha\ka'}\erpkp\erk
\eqn\eextratwo
$$
{}From eqs.\etwonea\ and \etwothreea\ we get,
$$\eqalign{
\tatw_{st}\atw_{t\alpha} &=-\l\Phi_{2,s}|c_0^-
Q_B|\Phi_{2,t}\r\l\Phi^c_{3,t}|
c_0^- Q_B|\cpa\r\cr
&=- \l\Phi_{2,s}| c_0^- Q_B b_0^- c_0^- Q_B|\cpa\r\cr
&= 0\cr
}
\eqn\efourthirteen
$$
In deriving the above equation, we have used the completeness relation
that follows from eq.\efourfive, and the nilpotence of the BRST charge
$(Q_B)^2=0$.
Thus the term involving $\kth$ in eq.\efourtwo\ also vanishes in this
case.
The equation then takes the form:
$$
\ctw_{\hat r i}\bth_{ij\ka}\btw_{j\ka'}\erpkp\erk
+\cth_{\hat rji}\btw_{i\ka}\btw_{j\ka'} \erpkp\erk =\ath_{\hat r\alpha
s}\dtw_{\alpha\ka} \ctw_{sj}\btw_{j\ka'}\erpkp\erk
\eqn\efourten
$$
{}From the definition of $\cth_{rji}$ given in eq.\etwofive\ we see that
it is symmetric under the exchange of $i$ and $j$.
(We have used this
symmetry property to get eq.\etwosix.)
Thus the term involving $\cth$ in eq.\efourten\ is symmetric under the
exchange of $\rho$ and $\rho'$.
If we exchange $\rho$ and $\rho'$ in eq.\efourten\ and subtract from the
original equation, we get,
$$
\ctw_{\hat r i}\bth_{ij\ka}\btw_{j\ka'}\erpkp\erk -
(\rho\leftrightarrow \rho')
= \ath_{\hat r\alpha s}\dtw_{\alpha\ka}\ctw_{sj}\btw_{j\ka'}\erpkp \erk
-(\rho\leftrightarrow \rho')
\eqn\efoureleven
$$
Note that all the undetermined constants $\cth$, $\dth$ and $\kth$ have
dropped out of the above equation, and hence it again represents a
possible obstruction to solving the set of equations given in
eq.\efourtwo.

Finally, let us choose $\erk$ in eq.\efourtwo\ to be of the form $\herk$
and multiply both sides of the equation by $\tatw_{rt'}\ctw_{t'j'}\pmpjp$.
In this case the term involving $\cth$ vanishes since
$\btw_{i\ka}\herk=0$.
The term involving $\dth$ is given by,
$$
\dth_{\alpha\ka j}\pmj\herk\atw_{r\alpha}\tatw_{rt'}\ctw_{t'j'}\pmpjp
\eqn\efourtwelve
$$
which vanishes by eq.\efourthirteen\ and the fact that $\tatw_{rs}$ is
symmetric in the indices $r$ and $s$.
Thus the equation takes the form:
$$\eqalign{
\ctw_{ri}\bth_{ij\ka}\pmj\herk\tatw_{rt'}\ctw_{t'j'}\pmpjp
&= \ath_{r\alpha s}\dtw_{\alpha\ka}\ctw_{sj}\pmj\herk\tatw_{rt'}\ctw_{t'j'}
\pmpjp\cr
&\qquad +\kth_{rs\ka}\tatw_{st}\ctw_{tj}\pmj\herk\tatw_{rt'}\ctw_{t'j'}\pmpjp
\cr
}
\eqn\efourfourteen
$$
As we have seen in sect.~2, the quantity $K_{rs}$ and hence $\kth_{rs\ka}$
defined in eq.\etwoeleven\ must be antisymmetric in $r$ and $s$.
Hence the term involving $\kth$ in eq.\efourfourteen\ in antisymmetric in
the indices $m$ and $m'$.
Thus if we symmetrize both sides of the equation in $m$ and $m'$, the term
involving $\kth$ drops out and we are left with the equation:
$$\eqalign{
&\ctw_{ri}\bth_{ij\ka}\pmj\herk\tatw_{rt'}\ctw_{t'j'}\pmpjp +
(m\leftrightarrow m')\cr
&\qquad\qquad= \ath_{r\alpha
s}\dtw_{\alpha\ka}\ctw_{sj}\pmj\herk\tatw_{rt'}\ctw_{t'j'} \pmpjp +
(m\leftrightarrow m')\cr
}
\eqn\efourfifteen
$$
Since the above equation involves only the known quantities, this
represents a third set of obstructions to solving the set of equations
\efourtwo.

We shall now show that once eqs.\efournine, \efoureleven\ and
\efourfifteen\ are satisfied, we can always find a solution of the set of
equations \efourone\ (or, equivalently, eqs.~\efourtwo). In other words
we can find
appropriate $\cth$, $\dth$ and $\kth$ satisfying these equations.
To start with, let us consider the case where the index $r$ is of type
$\tilde r$, i.e. $\l\tprc|Q_B\ne 0$.
In this case we may express eq.\efourone\ as,
$$
\atw_{\tilde r\alpha}\dth_{\alpha\ka j}=\ctw_{\tilde r i}\bth_{ij\ka}
+\cth_{\tilde r ji}\btw_{i\ka} -\ath_{\tilde r\alpha s}\dtw_{\alpha\ka}
\ctw_{sj} -\kth_{\tilde r s\ka}\tatw_{st}\ctw_{tj}
\eqn\efourthirtythree
$$
Let us now note that $\atw_{\tilde r\alpha}$ has no eigenvector with zero
eigenvalue acting on the left.
For, if there was such an eigenvector (say $x_{\tilde r}$) then it would
give,
$$
0=\sum_{\tilde r} x_{\tilde r}\atw_{\tilde r\alpha}=\sum_{\tilde r}
x_{\tilde r}\l\tprc|c_0^- Q_B|\cpa\r,\qquad\hbox{for all }\alpha
\eqn\efourthirtyfour
$$
which, in turn, would imply that $\sum_{\tilde r}x_{\tilde r}\l\tprc|$ is
annihilated by $Q_B$.
But we have already chosen the basis states such that all such states are
included in the set $\{\l\hprc|\}$; the set $\{\l\tprc|\}$ contains only
those states which are not annihilated by $Q_B$.
Thus there is no vector $x_{\tilde r}$ for which $\sum x_{\tilde
r}\atw_{\tilde r\alpha}=0$.
This, in turn, shows that the matrix $\atw_{\tilde r\alpha}$ has a
(non-unique) right
inverse $M_{\alpha\tilde s}$, satisfying,\foot{Although $\atw$ is an
infinite dimensional matrix, it is block diagonal in the basis where the
states are taken to be eigenstates of the momentum operator and $L_0^+$,
with each block being a finite matrix.
Thus all the results for finite dimensional matrices can be applied here.}
$$
\atw_{\tilde r\alpha}M_{\alpha\tilde s} =\delta_{\tilde r\tilde s}
\eqn\efourthirtyfour
$$
We thus get a solution of eq.\efourthirtythree\ of the form:
$$
\dth_{\alpha\ka j} =M_{\alpha\tilde r}[\ctw_{\tilde r i}\bth_{ij\ka} +
\cth_{\tilde r ji}\btw_{i\ka} -\ath_{\tilde r\alpha
s}\dtw_{\alpha\ka}\ctw_{sj} -\kth_{\tilde r s\ka}\tatw_{st}\ctw_{tj}]
\eqn\efourthirtyfive
$$
In other words, we can always adjust the coefficients $\dth_{\alpha\ka j}$
so as to satisfy eq.\efourthirtythree.

Next we consider the case where the index $r$ is taken to be of the type
$\hat r$, but the index $\rho$ (in eq.\efourtwo) is taken to be of the
type $\tilde\rho$.
In this case, the term involving $\dth$ in eq.\efourtwo\ vanishes, and
we may rewrite this equation as,
$$\eqalign{
&\cth_{\hat r ji}\btw_{i\ka}\pmj\terk
=(\ath_{\hat r\alpha s}\dtw_{\alpha\ka}\ctw_{sj} +\kth_{\hat r
s\ka}\tatw_{st} \ctw_{tj} -\ctw_{\hat r i}\bth_{ij\ka})\pmj\terk\cr
}
\eqn\efourthirtysix
$$
Let us consider the matrix $S_{i\tilde\rho}\equiv \btw_{i\ka}\terk$.
{}From our choice of basis described in eq.\efourfour\ it is clear that this
matrix cannot have an eigenvector with zero eigenvalue while acting on the
right; for if there was such an eigenvector (say $\{y_{\tilde\rho}\}$)
then the combination $y_{\tilde\rho}\terk$ would have to be included in
the set $\{\herk\}$, and the set $\{\herk\}$, $\{\terk\}$ will not
together form a set of {\it linearly independent} basis vectors.
This, in turn, shows that the matrix $S_{i\tilde\rho}$ must have a
(non-unique) left inverse $N_{\tilde\rho j}$ satisfying,
$$
N_{\tilde\rho i}\btw_{i\ka}\tilde\eta^{(\tilde\rho')}_\ka\equiv
N_{\tilde\rho i}S_{i\tilde\rho'} =\delta_{\tilde\rho\tilde\rho'}
\eqn\efourthirtyseven
$$
Thus we see that eq.\efourthirtysix\ is satisfied if we choose,
$$
\cth_{\hat r ji}= (\ath_{\hat r\alpha s}\dtw_{\alpha\ka}\ctw_{sj}
+\kth_{\hat r s\ka}\tatw_{st}\ctw_{tj} -\ctw_{\hat r l}\bth_{lj\ka})\terk
N_{\tilde\rho i}
\eqn\efourthirtyeight
$$
We must, however, make sure that the expression for $\cth_{\hat r ji}$
obtained this way is symmetric in $i$ and $j$.
To do this it is convenient to choose a specific basis for the variables
$\{\phi_i\}$, which makes the matrix $S_{i\tilde\rho}$ look simple.
Let us assume that $\tilde\rho$ takes $N$ different values.
(In general $N$ is infinite, but if we work within a subspace of states
with a given momentum, then $N$ is finite.)
Let us choose the basis for the variables $\phi_i$
such that\foot{Such a choice of basis is possible through a linear field
redefinition of the form $\phi_i\to W_{ij}\phi_j$, with
$$
W_{ij}=
\cases{N_{ij} &for $1\le i\le N$\cr
V_{ik}(\delta_{kj}-S_{k\tilde\rho'}N_{\tilde\rho' j}) &for $i>N$\cr
}
$$
where $V$ is a matrix which should be chosen so as to make $W$
non-singular, but is otherwise arbitrary.
}
$$
S_{i\tilde \rho} =
\cases{\delta_{i\tilde \rho} &for $1\le i\le N$\cr
0 &for $i>N$\cr
}
\eqn\etildesone
$$
In this case
$$
N_{\tilde\rho i} =
\cases{\delta_{\tilde\rho i} &for $1\le i\le N$\cr
arbitrary &for $i>N$\cr
}
\eqn\etildenone
$$
Hence eq.\efourthirtyeight\ may be written as,
$$
\cth_{\hat r ji}=
\cases{
(\ath_{\hat r\alpha
s}\dtw_{\alpha\ka}\ctw_{sj}+\kth_{\hat r s\ka}\tatw_{st}\ctw_{tj}
-\ctw_{\hat r l}\bth_{lj\ka})\tilde\eta^{(i)}_\ka, &for $1\le i\le
N$\cr
{\rm arbitrary} &for $i\ge N+1$.\cr
}
\eqn\efourthirtynine
$$
Thus we see that if $i\le N$ and $j>N$, then the relation $\cth_{\hat r
ij}=\cth_{\hat r ji}$ can be satisfied by setting $\cth_{\hat r ij}$
(which is undetermined otherwise) to be equal to $\cth_{\hat r ji}$.
For $i,j > N$ one can always choose $\cth_{\hat rij}$ to be symmetric in
$i$ and $j$, since it is not constrained at all.
Thus the only possible source of discrepancy comes when both $i$ and $j$
are less than or equal to $N$.
However, in this case, using eq.\etildesone\ we
may write,
$$
\cth_{\hat r ji} = \cth_{\hat r j' i} S_{j' j}, \qquad 1\le i, j\le N.
\eqn\efourforty
$$
Substituting the solution \efourthirtyeight\ on the right hand side of the
above equation, using the relation
$N_{\tilde\rho i}=\delta_{\tilde\rho i}$, and that $S_{j' j}
=\btw_{j' \ka}\tilde\eta^{(j)}_\ka$ for $1\le j\le N$, we get,
$$\eqalign{
\cth_{\hat r ji} = & (\ath_{\hat r\alpha s}\dtw_{\alpha\ka}\ctw_{sj'}
+\kth_{\hat r s\ka} \tatw_{st} \ctw_{tj'} -\ctw_{\hat r l}\bth_{lj'\ka} )
\tilde\eta^{(i)}_{\ka}\btw_{j'\ka '}\tilde\eta^{(j)}_{\ka'}\cr
&\qquad {\rm for}~1\le i,j\le N.\cr
}
\eqn\efourfortyone
$$
Since $\kth_{\hat r
s\ka}\tatw_{st}\ctw_{tj}\btw_{j\ka'}\tilde\eta^{(j)}_{\ka'}$
vanishes (see eqs.\eextratwo\ and \efourthirteen)
the term involving $\kth$ vanishes in the
above equation.
Eq.\efoureleven\ then implies that the right hand side of the equation is
symmetric in $i$ and $j$, showing that $\cth_{\hat r ji}$ calculated from
eq.\efourthirtyeight\ in indeed symmetric in $i$ and $j$.

Thus it remains to show that if in eq.\efourtwo\ we take the index $r$ to
be of type $\hat r$, and $\rho$ to be of the type $\hat\rho$, then this
equation can still be satisfied.
In this case the terms involving $\cth$ as well as $\dth$ drop out of the
equation.
The only other free index in this equation is $m$, which can either be of
type $\tilde m$, or of type $\hat m$.
If this index is of type $\hat m$ then eq.\efournine\ guarantees that
eq.\efourtwo\ is automatically satisfied.
Thus we now need to ensure that we can satisfy eq.\efourtwo\ by adjusting
$\kth$ when the indices $r$, $\rho$ and $m$ are of the type $\hat r$,
$\hat\rho$ and $\tilde m$ respectively.
In this case we can write eq.\efourtwo\ as,
$$
\kth_{\hat r s\ka}\tatw_{st}\ctw_{tj}\tpmj\herk
=\ctw_{\hat r i}\bth_{ij\ka}\tpmj\herk -\ath_{\hat r\alpha
s}\dtw_{\alpha\ka} \ctw_{sj}\tpmj\herk.
\eqn\efourfortytwo
$$
Let us define,
$$
T_{s\tilde m} =\tatw_{st}\ctw_{tj}\tpmj
\eqn\efourfortythree
$$
Using eq.\efourthree\ and arguments similar to the case of the matrix
$S_{i\tilde\rho}$, it is clear that $T_{s\tilde m}$, acting on the right,
does not have any eigenvector with zero eigenvalue.
As a result, it has a left inverse; let us call this $U_{\tilde m r}$:
$$
U_{\tilde m r}T_{r\tilde n} =\delta_{\tilde m\tilde n}
\eqn\efourfortyfour
$$
Also note that since $\{\eta^{(\rho)}\}$ form a complete set of linearly
independent vectors, $\erk$, regarded as a matrix in $(\rho,\ka)$ space,
is invertible.
Thus if we define,
$$
\ck^{(1)}_{rs\rho} =\kth_{rs\ka}\erk
\eqn\efourfortyfoura
$$
then we can freely obtain $K$ from $\ck$ and vice versa with the help of
the matrix $\erk$ or its inverse.
A solution to eq.\efourfortytwo\ is given by,
$$\eqalign{
\ck^{(1)}_{rs\hat\rho} &=
(\ctw_{ri}\bth_{ij\ka}-\ath_{r\alpha
s'}\dtw_{\alpha\ka} \ctw_{s' j})\tpmj U_{\tilde m s}\herk\cr
\ck^{(1)}_{rs\tilde\rho} &= \>{\rm arbitrary.}\cr
}
\eqn\efourfortyfive
$$
Note that only $\ck^{(1)}_{\hat r s\hat\rho}$ is determined by
eq.\efourfortytwo.
Eq.\efourfortyfive\ gives a specific solution of eq.\efourfortytwo, but
in general $\ck^{(1)}_{\tilde r s\hat\rho}$ can be chosen arbitrarily.
Finally we need to verify that $\ck^{(1)}$ determined this way is
antisymmetric in $r$ and $s$.
For this we assume that the index $\tilde m$ runs from 1 to $M$ and,
as in the case of the matrix $S_{i\tilde\rho}$, choose
a basis of states $|\Phi_{2,r}\r$ such that:
$$
T_{r\tilde m} =
\cases{\delta_{r\tilde m} &for $1\le r\le M$\cr
0 &for $r>M$\cr
}
\eqn\efourfortysix
$$
so that $U_{\tilde m r}$ can be taken to be,
$$
U_{\tilde m r} =
\cases{
\delta_{\tilde m r} &for $1\le r\le M$,\cr
{\rm arbitrary} &for $r>M$.\cr
}
\eqn\efourfortyseven
$$
In this basis $\ck^{(1)}_{rs\hat\rho}$ given in eq.\efourfortyfive\ takes
the form
$$
\ck^{(1)}_{rs\hat\rho} =
\cases{
(\ctw_{ri}\bth_{ij\ka} -\ath_{r\alpha
s'}\dtw_{\alpha\ka} \ctw_{s' j})\phi^{(s)}_j\herk, &for $1\le s\le M$;\cr
{\rm arbitrary} &for $s\ge M+1$.\cr
}
\eqn\efourfortyeight
$$
Thus we see that as long as either $s$ or $r$ is larger than $M$, the
antisymmetry of $\ck^{(1)}_{rs\hat \rho}$ may be satisfied by judicious
choice of $\ck^{(1)}_{rs\hat\rho}$ in the
region $s\ge M+1$, where it is undetermined otherwise.
The only possible problem comes from the range where both $r$ and $s$ lie
in the range between  1 and $M$.
But note that in this basis eq.\efourfifteen\ takes the form:
$$\eqalign{
&(\ctw_{ri}\bth_{ij\ka}\phi^{(s)}_j\herk -\ath_{r\alpha
s'}\dtw_{\alpha\ka} \ctw_{s' j} \phi^{(s)}_j\herk) + (r\leftrightarrow s)
=
0, ~~{\rm for}~1\le r,s\le M\cr
}
\eqn\efourfortynine
$$
which is precisely the statement that $\ck^{(1)}_{rs\hat\rho}$ given in
eq.\efourfortyeight\ is antisymmetric in $r$ and $s$.

This completes the proof that once eqs.\efournine, \efoureleven\ and
\efourfifteen\ are satisfied, we can choose appropriate $\cth$, $\dth$ and
$\kth$ so as to satisfy eq.\efourone.
We have shown explicitly in appendix A that these equations are indeed
satisfied.
This in turn means that it is possible to identify appropriate string
field configurations to off shell field configurations in low energy
effective field theory, and to identify appropriate symmetries in string
field theory to gauge symmetries in low energy effective field theory, so
as to get the correct transformation laws of various fields in low energy
effective field theory to first non-linear order.

Before we conclude this section, we would like to show that a solution to
eq.\efourone\ (or eq.\efourtwo) cannot be obtained if we set
$\kth_{rs\ka}=0$.
For this let us take $\kth_{rs\ka}=0$, and choose $\erk$, and $\l\prc|$ to
be of the forms $\herk$ and $\l\hprc|$ respectively.
As a result, terms involving $\cth$ and $\dth$ drop out of eq.\efourtwo\
and it takes the form:
$$
\ctw_{\hat r i}\bth_{ij\ka}\pmj\herk =\ath_{\hat r\alpha
s}\dtw_{\alpha\ka} \ctw_{sj}\pmj\herk
\eqn\eextrafour
$$
Since this equation involves only known coefficients, this is a
consistency equation.
In appendix A we show one example of a specific choice of $\l\hprc|$,
$\herk$ and $\pmj$ for which this equation is not satisfied.
This, in turn, shows that it is not possible to obtain solutions of
eq.\efourtwo\ with $\kth=0$.

\chapter{DISCUSSION}

In this paper we have studied how off-shell general coordinate
transformations and antisymmetric tensor gauge transformations arise in
string field theory.
Working to first non-linear order, we have shown that it is possible to
identify specific string field configurations with off-shell field
configurations in low energy effective field theory, and specific symmetry
transformations in string theory with off-shell gauge transformations in
low energy effective field theory, so that the symmetry transformations in
the former theory are compatible with those in the latter theory.

One of the specific results of our analysis is that the off-shell gauge
symmetries of low energy effective field theory cannot be identified to
just a combination of off-shell gauge symmetries alone of string field theory.
Instead, they can be identified to a combination of off-shell gauge
symmetries of string field theory and the trivial symmetries of the form
given in eq.~\eoneone.

The gauge algebra of string field theory is characterized by two important
features.
The first is that it closes only on-shell; the second is that the
algebra has field dependent structure constants.
More specifically, if $|\Lambda_1\r$ and $|\Lambda_2\r$ are two
independent gauge transformation parameters, the commutator of these
two gauge transformation parameters, acting on the string field $|\Psi\r$
gives\foot{This observation has been made independently by
Zwiebach\ZWIEPRI}:
$$
[\delta_{\Lambda_1},\delta_{\Lambda_2}]\psi_r
=\delta_{\Lambda} \psi_r + M_{rs}(\{\psi_t\}){\delta S\over \delta\psi_s}
\eqn\econctwo
$$
where,
$$
b_0^-|\Lambda\r = \sum_{N=0}^\infty {g^{N+1}\over N!}
[\Lambda_2\Lambda_1 \Psi^{N}]
\eqn\econctwo
$$
and $M_{rs}$ is an antisymmetric matrix, given by,
$$
M_{rs}(\Psi,\Lambda_1, \Lambda_2) = \sum_{N=0}^\infty {g^{N+2}\over N!}
\{(c_0^-\Phi^c_{3,s}) (c_0^-\Phi^c_{3,r}) \Psi^N\Lambda_1\Lambda_2\}
\eqn\econclatest
$$
Antisymmetry of $M_{rs}$ follows from the property of
$\{~~\}$\GAUGEINV\SENPOLY.
On the other hand, the gauge symmetries of the low energy effective
Lagrangian closes off-shell, and have field independent structure
constants.
Thus if our analysis can be extended to all orders in the string
field $\Psi$, this would imply that the string field theory contains
a symmetry (sub-)algebra which does close even when the massless field
in the theory are off-shell (whereas the massless auxiliary fields and the
massive fields are eliminated by
their equations of motion).
It is therefore natural to ask how to reconcile these apparently
different features of string
field theory and low energy effective field theory.

Let us first address the question of closure of the algebra.
Note that although the commutator of two gauge transformations in string
field theory contains
trivial symmetry transformations given in eq.\eoneone, and hence the gauge
algebra closes only on-shell, it is conceivable
that one can
redefine the gauge transformation laws by adding appropriate combination
of the
trivial symmetry transformations to each gauge transformation, so that the
resulting algebra (or some subalgebra of the resulting algebra) closes
off-shell.
We expect that this is precisely what happens in this case.
In fact,
from our analysis, we have already seen that the gauge transformations of
low
energy effective field theory indeed correspond to  combinations of gauge
transformation and the trivial symmetry in string field theory.

On the other hand, the structure constants of the algebra can be changed
by appropriate redefinition of gauge transformation parameters.
(A somewhat contorted example is of a $U(1)$ gauge theory, where we could
have defined the gauge
transformation law of the gauge field $A_\mu$ to $\delta A_\mu
=\p_\mu((1+f(A))\eps)$ where $f$ is some function of $A_\mu$.
This would, in general, give field dependent structure constants for the
gauge group, although the standard $U(1)$ algebra has field
independent structure constants.)
Thus it is not surprising that one can obtain suitable (field dependent)
combination of gauge transformations in string field theory to get a
subalgebra of the gauge group
with field independent structure constants.
To show that such combinations can really be obtained we need to extend
the analysis of the paper to higher orders in the fields.

We expect that the extension of our analysis to higher orders in $\Psi$
can be carried out using a method of induction, where we assume that a
solution of the set of first $N$ equations appearing in eq.\etwofourteen\
have been obtained, and then prove that the $(N+1)$th equation in that set
can also be solved.
We hope to come back to this question in the future.

\Appendix{A}

\centerline{\bf EXPLICIT VERIFICATION OF THE CONSISTENCY CONDITIONS}

In this appendix we shall show that the consistency conditions represented
by eqs.\efournine, \efoureleven\ and \efourfifteen\ are indeed satisfied
by the vertices of string field theory. We also demonstrate by one example
that eq.~\eextrafour\ is in general not satisfied.

We begin with eq.\efournine. Let us define,
$$
b_0^-|\hpsm\r =\ctw_{sj}\hpmj |\Phi_{2,s}\r
\eqn\efoursixteen
$$
and,
$$
b_0^-|\hlr\r = \dtw_{\alpha\ka}\herk |\cpa\r
\eqn\efourseventeen
$$
{}From this we see that,
$$
\l\Phi_{2,r}|c_0^- Q_B b_0^-|\hpsm\r =-\ctw_{sj}\hpmj\tatw_{rs} =0
\eqn\efoureighteen
$$
and,
$$
\l\Phi^c_{3,r}| c_0^- Q_B b_0^- |\hlr\r = \dtw_{\alpha\ka}\herk\atw_{r\alpha}
=\ctw_{ri} \btw_{i\ka}\herk = 0
\eqn\efournineteen
$$
using eq.\efourthree\ and eqs.\ethreetwelvea\ and \efourfour\
respectively.
This, in turn, shows that $Q_B b_0^-|\hpsm\r$ and $Q_B b_0^-|\hlr\r$
vanish identically.

Using eq.\etwothreea\ and the definition of $\{~~\}$\GAUGEINV\SENPOLY\ we
get
$$
\ath_{\hat r\alpha s} =
g \l f_1\c \hprc(0) f_2\c \cpa(0) f_3\c\Phi_{2,s}(0)\r
\eqn\efourtwenty
$$
where $f_i$ are known conformal maps\LPP\NONPOL\GAUGEINV\SENPOLY.
{}From this we see that the right hand side
of eq.\efournine\ takes the form:
$$
g\l f_1\c \hprc(0) f_2\c b_0^-\hlr(0) f_3\c b_0^-\hpsm(0)\r
\eqn\efourtwentyone
$$
where $b_0^-\hlr$ and $b_0^-\hpsm$ are the local fields in conformal field
theory
which create the states $b_0^-|\hlr\r$ and $b_0^-|\hpsm\r$ acting on the
vacuum\BPZ.
As we have seen, $\hprc$, $b_0^-\hlr$ and $b_0^-\hpsm$ are all BRST
invariant fields.
Hence if any of them is a BRST trivial field, then expression
\efourtwentyone\ vanishes identically, $-$ the only contribution comes
from the term when $\hprc$, $\hlr$ and $\hpsm$ all correspond to
(non-zero) elements of BRST cohomology.

We thus first need to verify that the left hand side of eq.\efournine\
vanishes when either of $\hprc$, $\hlr$ or $\hpsm$ is BRST trivial.
To this end, note that  under the antisymmetric tensor gauge
transformation, the transformation of the fields have only linear term
($\delta B_{\mu\nu}\propto \p_\mu\xi_\nu-\p_\nu\xi_\mu$), hence
$\bth_{ij\ka}$ vanishes in this case.
Thus we need to look for contribution to the left hand side of
eq.\efournine\ from general coordinate transformation.
For a general coordinate transformation labelled by the parameter
$\eps_\mu(x)$, we see from eq.\ethreeeleven\ that,
$$
b_0^-|\hlr\r =-i\int d^D k\heps_\mu(k) (c_1\alpha^\mu_{-1} -\bar
c_1\bar\alpha^\mu_{-1}) |k\r
\eqn\efourtwentytwo
$$
It is easy to see that for no $\heps_\mu$ this can be written as
$Q_B|s\r$.
Thus $b_0^-|\hlr\r$ must be a (non-zero) member of the BRST
cohomology.
A straightforward analysis of the BRST cohomology shows that such states
are given by $\heps_\mu(k)=-(1/\sqrt 2 g)\eps_\mu\delta^{(D)}(k)$,
corresponding to rigid translation by an amount $\eps_\mu$ (see
eqs.\ethreenine).
Thus in this case,
$$
\bth_{ij\ka}\hpmj\herk = i\eps_\mu k^\mu\hat\phi^{(\hat m)}_i
\eqn\eextraone
$$
where $k$ is the $D$-momentum carried by $\hpmj$.\foot{We have chosen a
basis $\{\pmj\}$ such that $\pmj$ has a fixed momentum.}
Since $\ctw_{ri}$ is block diagonal in the momentum space
(i.e. if $\pmj$ carries momentum $k$ then $b_0^- |\Psi^{(m)}\r\equiv
\ctw_{r j}\pmj |\Phi_{2,r}\r$ also carries momentum $k$), we can express
$\ctw_{\hat r i}\bth_{ij\ka}\hpmj$ as $(i\eps_\mu k^\mu)\ctw_{\hat r j}
\hpmj$.
Using eq.\efoursixteen\ we now see that,
$$
\ctw_{\hat r j}\hpmj =\l\hprc |\hpsm\r
\eqn\efourtwentythree
$$
Hence if either $\l\hprc|$ or $b_0^-|\hpsm\r$ is BRST trivial, $\ctw_{\hat
r j}\hpmj$ vanishes.
This, in turn, shows that the left hand side of eq.\efournine\ also
vanishes unless $\l\hprc|$, $b_0^-|\hpsm\r$ and $b_0^-|\hlr\r$ are all
(non-zero) elements of the BRST cohomology.

We can now restrict our attention to the case where
$\l\hprc|$, $|\hpsm\r$ and $|\hlr\r$ are
all (non-zero) elements of the BRST cohomology.
First let us consider the case where $|\hlr\r$ corresponds to gauge
transformation associated with antisymmetric tensor field.
As remarked before, in this case the left hand side of eq.\efournine\
vanishes.
The right hand side is given by eq.\efourtwentyone.
Standard analysis of BRST cohomology shows that $b_0^-|\hlr\r$
must have the form $\xi_\mu (c_1\alpha^\mu_{-1}+\bar
c_1\bar\alpha^\mu_{-1})|0\r$, whereas $b_0^-|\hpsm\r$ and $\l\hprc|$ can be
taken to be of the form $a_{\mu\nu}(k)c_1\bar
c_1\alpha^\mu_{-1}\bar\alpha^\nu_{-1}|k\r$, and
$\l -k|\alpha^\mu_1\bar\alpha^\nu_1 \bar c_{-1}c_{-1}c_0^+e_{\mu\nu}(k)$
respectively, with
$k^\mu a_{\mu\nu}=k^\nu a_{\mu\nu}=k^\mu e_{\mu\nu}=k^\nu e_{\mu\nu} =0$,
$k^2=0$.
Since each of these are primary states, it is straightforward to compute
expression \efourtwentyone.
It turns out to vanish, showing that eq.\efournine\ is satisfied for the case
where $\herk$ is taken to be an antisymmetric tensor gauge transformation.

We now look at the case where $|\hlr\r$ (or equivalently $\herk$)
correspond to general coordinate transformation.
As remarked before, the only (non-zero) element of BRST cohomology is
generated by rigid translation, and using eq.\efourtwentythree\ and the
discussion above it, we can bring the left hand side of eq.\efournine\ to
the form:
$$
i\eps_\mu k^\mu\l\prc |\hpsm\r
\eqn\efourtwentyfour
$$
The right hand side of eq.\efournine\ is given by eq.\efourtwentyone.
{}From eq.\efourtwentytwo\ and the fact that $\hat \eps_\mu =  (-1/\sqrt 2
g) \eps_\mu\delta^{(D)}(k)$ (eq.\ethreenine), we get,
$$
b_0^-|\hlr\r = {i\over \sqrt 2 g} \eps_\mu (c_1\alpha^\mu_{-1} -\bar
c_1\bar\alpha^\mu_{-1}) |0\r
\eqn\efourtwentyfoura
$$
$\l\hprc|$ and $b_0^-|\hpsm\r$ have the same form as given in the previous
paragraph.
Since $\hprc$, $b_0^-\hlr$ and $b_0^-\hpsm$ are all dimension $(0,0)$
primary fields, it is straightforward to evaluate eq.\efourtwentyone\ and
we get the answer:
$$
i\eps_\mu k^\mu \l\hprc |\hpsm\r
\eqn\efourtwentyfive
$$
which agrees with eq.\efourtwentyfour.

Finally, note that
if $k=0$, then the set $\{|\hpsm\r\}$ contains an extra physical state of
the form $(c_1 c_{-1} -\bar c_1\bar c_{-1})|0\r$, and the set
$\{\l\hprc|\}$ contains an extra physical state of the form $\l 0|(c_1
c_{-1} -\bar c_1\bar c_{-1}) c_0^+$.
It is easy to see that both for $|\hlr\r$ representing a rigid translation
or a rigid antisymmetric tensor gauge transformation, the right and the
left hand side of eq.\efournine\ vanishes.

This completes the proof  that eq.\efournine\ is satisfied for all values
of $\hat r$, $\hat\rho$ and $\hat m$.

We now turn to eq.\efoureleven.
Defining $b_0^-|\Lambda^{(\rho)}\r=\erk
\dtw_{\alpha\ka}|\cpa\r$ as before, and noting that,
$$\eqalign{
\ctw_{sj}\btw_{j\ka'}\erpkp |\Phi_{2,s}\r = & \atw_{s\alpha}\dtw_{\alpha\ka'}
\erpkp |\Phi_{2,s}\r\cr
= & \l\Phi_{3,s}^c|c_0^- Q_B|\cpa\r \dtw_{\alpha\ka'}\erpkp
|\Phi_{2,s}\r\cr
= & Q_B b_0^- |\lrp\r\cr
}
\eqn\efourtwentysix
$$
(we have used eq.\ethreetwelvea); we can express the right hand side of
eq.\efoureleven\ as,
$$
g[\l f_1\c \hprc(0) f_2\c b_0^-\lr(0) f_3\c Q_B b_0^-\lrp(0)\r
-\l f_1\c\hprc(0) f_2\c \lrp(0) f_3\c Q_Bb_0^-\lr(0)\r]
\eqn\efourtwentyseven
$$
Since the closed string vertex is completely symmetric, we can replace
$f_2$ by $f_3$ and $f_3$ by $f_2$ in the last term in
eq.\efourtwentyseven.
Using the fact that $\hprc$ is BRST invariant, we can deform the $Q_B$
contour in the first term so that it acts on $b_0^-\lr(0)$ instead of
$b_0^-\lrp(0)$.
We pick up two minus signs in this process, one from reversing the
contour, the other from commuting $Q_B$ through $b_0^-\lr$.
The result is that the first term exactly cancels the second term, showing
that the right hand side of eq.\efoureleven\ vanishes identically.

What about the left hand side?
Note that $\phi_j \equiv \btw_{j\ka}\erk$ represents a field
configuration of the low energy effective field theory that is obtained
from the zero field configuration by the gauge transformation generated by
$\erk$.
Acting on this field configuration with the gauge transformation $\erpkp$
we generate a field configuration,
$$
\phi_i+\btw_{i\ka'}\erpkp+\bth_{ij\ka'}\phi_j\erpkp
= \btw_{i\ka}\erk +\btw_{i\ka'}\erpkp +\bth_{ij\ka'}\erpkp \btw_{j\ka}
\erk
\eqn\efourtwentyeight
$$
Antisymmetrizing the above expression in $\rho$ and $\rho'$, we get a
field configuration generated by the commutator of the two transformations
generated by $\erk$ and $\erpkp$.
This configuration is given by,
$$
\bth_{ij\ka}\btw_{j\ka'}\erk\erpkp - (\rho\leftrightarrow \rho')
\eqn\efourtwentynine
$$
Since the gauge algebra of the low energy effective field theory closes
off-shell, the commutator of two gauge transformations is another gauge
transformation.
Let $\eta^{(0)}_\ka$ be the parameter labelling this new gauge
transformation.
Then eq.\efourtwentynine\ may be expressed as,
$$
\btw_{i\ka}\eta^{(0)}_\ka
\eqn\efourthirty
$$
Using eq.\ethreetwelvea, and the fact that
$\atw_{\hat r\alpha} =\l\hprc|c_0^- Q_B|\cpa\r= 0$,
the left hand side of eq.\efoureleven\ may be written as,
$$
\ctw_{\hat r i}\btw_{i\ka}\eta^{(0)}_\ka =\atw_{\hat
r\alpha}\dtw_{\alpha\ka} \eta^{(0)}_\ka =0
\eqn\efourthirtyone
$$
Thus we see that the left hand side of eq.\efoureleven\ also vanishes
identically.
This proves that eq.\efoureleven\ is satisfied for all values of $\hat r$,
$\rho$ and $\rho'$.

Let us now consider eq.\efourfifteen.
If we define $b_0^-|\psmp\r=\ctw_{t' j'}\pmpjp |\Phi_{2,t'}\r$
and $b_0^-|\hlr\r=\herk \dtw_{\alpha\ka}|\cpa\r$, then, by
standard manipulations, the right hand side of eq.\efourfifteen\ may be
shown to be proportional to,
$$
[\l f_1\c Q_B b_0^-\psmp(0) f_2\c b_0^-\hlr(0) f_3\c b_0^-\Psi^{(m)}(0)\r +
(m\leftrightarrow m')]
\eqn\efourthirtytwo
$$
Since $Q_B b_0^-|\hlr\r =0$, we can again deform the BRST contour in the
first term so that $Q_B$ acts on $b_0^-\Psi^{(m)}(0)$.
Using the symmetry of the vertex we can also interchange $f_1$ and $f_3$
in the second term, and make appropriate rearrangement of the operators
inside the correlator, picking up appropriate signs in the process.
The final result is that the two terms in eq.\efourthirtytwo\ exactly
cancel each other, thereby showing that the right hand side of
eq.\efourfifteen\ vanishes identically.

It thus remains to show that the left hand side of eq.\efourfifteen\ also
vanishes identically.
To see this, let us write the action for the low energy effective
field theory in the following form:
$$
S_{eff}(\phi) =\sum_{N=2}^\infty {1\over N}\tilde B^{(N)}_{i_1\ldots i_N}
\phi_{i_1}\ldots \phi_{i_N}
\eqn\eaddone
$$
Invariance of this action under the gauge transformation given in
eq.\etwofour\ then gives,
$$\eqalign{
\tilde\btw_{ij}\btw_{j\ka} =&0\cr
\tilde\bth_{ijl}\btw_{i\ka}+\tilde\btw_{ij}\bth_{il\ka} +(j\leftrightarrow
l)= & 0\cr
}
\eqn\eaddtwo
$$
Multiplying the second of eq.\eaddtwo\ by $\herk$ and using eq.\efourfour\
we get,
$$
\tilde\btw_{ij}\bth_{il\ka}\herk + (j\leftrightarrow l) =0
\eqn\eaddthree
$$
Using eqs.\etwoone, \etwofive, and \eaddone\ we get,
$$
\tilde\btw_{ij} =\tatw_{rt}\ctw_{ri}\ctw_{tj}
\eqn\eaddfour
$$
(Note that this relation has already been verified in sect.~3, where the
quadratic part of the action obtained from string field theory was shown
to agree with that of low energy effective field theory.)
Eq.\eaddthree\ then takes the form,
$$
\tatw_{rt}\ctw_{ri}\ctw_{tj}\bth_{il\ka}\herk + (j\leftrightarrow l) =0
\eqn\eaddfive
$$
Vanishing of the left hand side of eq.\efourfifteen\ is an immediate
consequence of this equation.

Hence both sides of eq.\efourfifteen\ vanish, and are, therefore equal to
each other.
This completes the verification of eqs.\efournine, \efoureleven\ and
\efourfifteen.

Finally we give an example to show that eq.\eextrafour\ breaks down for
specific choices of $\hprc$, $\herk$ and $\pmj$.
We choose,
$$
b_0^-|\hlr\r \equiv \dtw_{\alpha\ka}\herk|\cpa\r = {i\over\sqrt 2
g}\eps_\mu
(c_1\alpha^\mu_{-1} -\bar c_1\bar\alpha^\mu_{-1})|0\r
\eqn\eextrafoura
$$
$$
\l\hprc|= a_{\mu\nu}\l -k|c_{-1}\bar c_{-1}\alpha^\mu_1\bar\alpha^\nu_1
Q_B
\eqn\eextrafourb
$$
and,
$$\eqalign{
b_0^-|\psmj\r\equiv & \ctw_{rj}\pmj |\Phi_{2,r}\r\cr
=& [h_{\mu\nu} c_1\bar c_1\alpha^\mu_{-1}\bar\alpha^\nu_{-1} -{1\over\sqrt
2} (k^\nu h_{\nu\mu}-\half k_\mu h_\nu^\nu) c_0^+ (c_1\alpha^\mu_{-1}-\bar
c_1\bar\alpha^\mu_{-1}) \cr
&\qquad +\half h_\mu^\mu (c_1c_{-1}-\bar c_1\bar c_{-1})]|k\r\cr
}
\eqn\eextrafivea
$$
with $h_{\mu\nu}=h_{\nu\mu}$.
In other words, $\herk$ denotes a rigid translation with parameter
$\eps_\mu$, $\l\hprc|$ denotes a
pure gauge (BRST-exact) state and $\pmj$
corresponds to an off-shell graviton background with momentum $k$ (see
eq.\ethreetwelve).
In this case the left hand side of the equation is proportional to,
$$
i\eps\cdot k \ctw_{\hat r j}\pmj = i\eps.k \l\hprc
|\psmj\r
\eqn\eextrafiveb
$$
On the other hand, the right hand side of the equation is given by,
$$
g\l f_1\c\hprc(0) f_2\c b_0^-\hlr(0) f_3\c b_0^-\psmj(0)\r
\eqn\eextrasix
$$
which can be shown to be equal to
$$
i\eps.k \l\hprc |\psmj\r + {\rm ~extra ~terms ~proportional ~to ~the~
equations~ of~ motion.}
\eqn\eextraseven
$$
The expressions \eextrafiveb\ and \eextrasix\ therefore differ
unless the background fields $\pmj$ are
solutions of the equations of motion.
This, in turn, shows that we cannot solve the set of equations \efourone\
with $\kth=0$.

\refout
\end